\documentclass[12pt,epsf]{article}
\usepackage{amsmath}
\usepackage{amssymb}
\begin{document}
%%%%%%%%%%%%%%%%%  Defs. %%%%%%%%%%%%%%%%%%%%%%%%%%%%%%
\def\a{\alpha}
\def\b{\beta}
\def\c{\varepsilon}
\def\d{\delta}
\def\e{\epsilon}
\def\f{\phi}
\def\g{\gamma}
\def\h{\theta}
\def\k{\kappa}
\def\l{\lambda}
\def\m{\mu}
\def\n{\nu}
\def\p{\psi}
\def\q{\partial}
\def\r{\rho}
\def\s{\sigma}
\def\t{\tau}
\def\u{\upsilon}
\def\v{\varphi}
\def\w{\omega}
\def\x{\xi}
\def\y{\eta}
\def\z{\zeta}
\def\D{{\mit \Delta}}
\def\G{\Gamma}
\def\H{\Theta}
\def\L{\Lambda}
\def\F{\Phi}
\def\P{\Psi}

\def\S{\Sigma}

\def\o{\over}
\def\beq{\begin{eqnarray}}
\def\eeq{\end{eqnarray}}
\newcommand{\gsim}{ \mathop{}_{\textstyle \sim}^{\textstyle >} }
\newcommand{\lsim}{ \mathop{}_{\textstyle \sim}^{\textstyle <} }
\newcommand{\vev}[1]{ \left\langle {#1} \right\rangle }
\newcommand{\bra}[1]{ \langle {#1} | }
\newcommand{\ket}[1]{ | {#1} \rangle }
\newcommand{\EV}{ {\rm eV} }
\newcommand{\KEV}{ {\rm keV} }
\newcommand{\MEV}{ {\rm MeV} }
\newcommand{\GEV}{ {\rm GeV} }
\newcommand{\TEV}{ {\rm TeV} }
\def\diag{\mathop{\rm diag}\nolimits}
\def\Spin{\mathop{\rm Spin}}
\def\SO{\mathop{\rm SO}}
\def\O{\mathop{\rm O}}
\def\SU{\mathop{\rm SU}}
\def\U{\mathop{\rm U}}
\def\Sp{\mathop{\rm Sp}}
\def\SL{\mathop{\rm SL}}
\def\tr{\mathop{\rm tr}}

\def\IJMP{Int.~J.~Mod.~Phys. }
\def\MPL{Mod.~Phys.~Lett. }
\def\NP{Nucl.~Phys. }
\def\PL{Phys.~Lett. }
\def\PR{Phys.~Rev. }
\def\PRL{Phys.~Rev.~Lett. }
\def\PTP{Prog.~Theor.~Phys. }
\def\ZP{Z.~Phys. }

% draw box with width #1 pt and line thickness #2 pt
\newcommand{\drawsquare}[2]{\hbox{%
\rule{#2pt}{#1pt}\hskip-#2pt%  left vertical
\rule{#1pt}{#2pt}\hskip-#1pt%  lower horizontal
\rule[#1pt]{#1pt}{#2pt}}\rule[#1pt]{#2pt}{#2pt}\hskip-#2pt%  upper horizontal
\rule{#2pt}{#1pt}}% right vertical

\def\vbr{\vphantom{\sqrt{F_e^i}}}% vertical brace for tables
% Young tableaux
\newcommand{\fund}{\drawsquare{6.5}{0.4}}%  fundamental
\newcommand{\afund}{\overline{\fund}}
\newcommand{\symm}{\drawsquare{6.5}{0.4}\hskip-0.4pt%
        \drawsquare{6.5}{0.4}}%  symmetric second rank tensor
\newcommand{\asymm}{\raisebox{-3pt}{\drawsquare{6.5}{0.4}\hskip-6.9pt%
        \raisebox{6.5pt}{\drawsquare{6.5}{0.4}}}}%  antisymmetric second rank
\newcommand{\asymmthree}{\raisebox{-7pt}{\drawsquare{6.5}{0.4}}\hskip-6.9pt%
\raisebox{-0.5pt}{\drawsquare{6.5}{0.4}}\hskip-6.9pt%
\raisebox{6pt}{\drawsquare{6.5}{0.4}}}% antisymmetric third rank
\newcommand{\asymmfour}{\raisebox{-10pt}{\drawsquare{6.5}{0.4}}\hskip-6.9pt%
\raisebox{-3.5pt}{\drawsquare{6.5}{0.4}}\hskip-6.9pt%
\raisebox{3pt}{\drawsquare{6.5}{0.4}}\hskip-6.9pt%
        \raisebox{9.5pt}{\drawsquare{6.5}{0.4}}}%  antisymmetric fourth rank
\newcommand{\Ythrees}{\raisebox{-.5pt}{\drawsquare{6.5}{0.4}}\hskip-0.4pt%
          \raisebox{-.5pt}{\drawsquare{6.5}{0.4}}\hskip-0.4pt% 
          \raisebox{-.5pt}{\drawsquare{6.5}{0.4}}}%  symmetric third rank
\newcommand{\Yfours}{\raisebox{-.5pt}{\drawsquare{6.5}{0.4}}\hskip-0.4pt%
          \raisebox{-.5pt}{\drawsquare{6.5}{0.4}}\hskip-0.4pt% 
          \raisebox{-.5pt}{\drawsquare{6.5}{0.4}}\hskip-0.4pt% 
          \raisebox{-.5pt}{\drawsquare{6.5}{0.4}}}%  symmetric fourth rank
\newcommand{\Ythreea}{\raisebox{-3.5pt}{\drawsquare{6.5}{0.4}}\hskip-6.9pt%
        \raisebox{3pt}{\drawsquare{6.5}{0.4}}\hskip-6.9pt
        \raisebox{9.5pt}{\drawsquare{6.5}{0.4}}}
\newcommand{\Yfoura}{\raisebox{-3.5pt}{\drawsquare{6.5}{0.4}}\hskip-6.9pt%
        \raisebox{3pt}{\drawsquare{6.5}{0.4}}\hskip-6.9pt
        \raisebox{9.5pt}{\drawsquare{6.5}{0.4}}\hskip-6.9pt
        \raisebox{16pt}{\drawsquare{6.5}{0.4}}}
\newcommand{\Yadjoint}{\raisebox{-3.5pt}{\drawsquare{6.5}{0.4}}\hskip-6.9pt%
        \raisebox{3pt}{\drawsquare{6.5}{0.4}}\hskip-0.4pt
        \raisebox{3pt}{\drawsquare{6.5}{0.4}}}%  SU(3) adjoint
\newcommand{\Ysquare}{\raisebox{-3.5pt}{\drawsquare{6.5}{0.4}}\hskip-0.4pt%
        \raisebox{-3.5pt}{\drawsquare{6.5}{0.4}}\hskip-13.4pt%
        \raisebox{3pt}{\drawsquare{6.5}{0.4}}\hskip-0.4pt%
        \raisebox{3pt}{\drawsquare{6.5}{0.4}}}%  4 boxes in a square
\newcommand{\Yflavor}{\Yfund + \overline{\Yfund}} % box anti-box pair
\newcommand{\Yoneoone}{\raisebox{-3.5pt}{\drawsquare{6.5}{0.4}}\hskip-6.9pt%
        \raisebox{3pt}{\drawsquare{6.5}{0.4}}\hskip-6.9pt%
        \raisebox{9.5pt}{\drawsquare{6.5}{0.4}}\hskip-0.4pt%
        \raisebox{9.5pt}{\drawsquare{6.5}{0.4}}}%

%%%%%%%%%%%%%%%%%%%%%%%%%%%%%%%%%%%%%%%%%%%%%%%%%%%
%%%%%%%%%%%%%%%%%%%%%%%%%%%%%%%%%%%%%%%%%%%%%%%%%%%
\baselineskip 0.7cm

\begin{titlepage}

\begin{flushright}
IPMU 08-0019\\
SLAC-PUB-13195\\
UCB-PTH-08/08
\end{flushright}

\vskip 1.35cm
\begin{center}
{\large \bf
Conformal Gauge Mediation \\ and Light Gravitino of Mass $m_{3/2} <O(10) ~\rm{eV}$ }
\vskip 1.2cm
M. Ibe${}^{1}$, Y. Nakayama${}^{2}$ and T.T. Yanagida${}^{3,4}$

\vskip 0.4cm
${}^{1}$ {\it Stanford Linear Accelerator Center, Stanford University,
                Stanford, CA 94309 and} \\
{\it Physics Department, Stanford University, Stanford, CA 94305}\\
${}^{2}${\it Berkeley Center for Theoretical Physics and Department of Physics,
\\
University of California, Berkeley, California 94720-7300}

${}^3${\it Department of Physics, University of Tokyo,     Tokyo 113-0033, Japan}

${}^4${\it Institute for the Physics and Mathematics of the Universe, University of Tokyo, \\
Kashiwa 277-8568, Japan}

\vskip 1.5cm

\abstract{We discuss a class of gauge mediated supersymmetry breaking models with conformal invariance above the messenger mass scale (conformal gauge mediation). The spectrum of the supersymmetric particles including the gravitino is uniquely determined by the messenger mass.
When the conformal fixed point is strongly interacting, 
it predicts a light gravitino of mass $m_{3/2}<O(10)$\,eV, which is attractive since such a light gravitino causes no problem in cosmology. 
}
\end{center}
\end{titlepage}

\setcounter{page}{2}

%%%%%%%%%%%%%%%%%%%%%%%%%%%%%%%%%%%%%%%%%%%%%%%%%%%%%%%%%%%%%%%%%%%%%%%%%%%%%%%%%%%%%%%%%%%%%%%%%%%%%%
\section{Introduction}

``Conformal gauge mediation" proposed in \cite{Ibe:2007wp} is a novel class of gauge mediation of supersymmetry (SUSY) breaking with strong predictive power. 
In the conventional gauge mediation
models~\cite{Dine:1981za}--\cite{Dine:1995ag}, 
although the gaugino and sfermion masses are related,
the gravitino mass is essentially a free-parameter that depends on the detail of the messenger couplings. In contrast, an advantage of the conformal gauge mediation is that the spectrum only depends on the conformal breaking scale, and, in particular, the gravitino mass is completely fixed by the SUSY breaking dynamics, enhancing our low-energy predictability.

The fundamental reason why we obtain this strong predictability in the conformal gauge mediation is due to the conformal invariance near the cut-off scale where the theory is defined. The assumption of the conformal invariance fixes the coupling constants of the SUSY breaking sector at their fixed point values, and they do not take arbitrary values in the low energy prediction. The only relevant deformation --- mass of the messengers in our construction, will yield the scale of the theory, determining the messenger scale, the conformal breaking scale and eventually the SUSY breaking scale as well.

This ``uniqueness" of the theory leads us to the analogy~\cite{Ibe:2007wp}  between QCD and the conformal gauge mediation. QCD, in the massless quark limit, is a marvelous unification of the Hadron physics in that the low energy predictions only depend on the QCD scale. Similarly, the conformal gauge mediation unifies the dynamics of the SUSY breaking and the messenger physics so that the low energy predictions only depend on the conformal breaking scale.

In this paper, as announced in~\cite{Ibe:2007wp}, we further examine strongly interacting examples of the conformal gauge mediation. We show that the requirement to avoid the splitting SUSY spectrum naturally gives rise to the strongly interacting conformal gauge mediation. Surprisingly, the strongly interacting conformal gauge mediation reveals an attractive feature from the cosmological viewpoint. In this model, the gravitino mass is as small as $O(1)$~eV, 
in which case there is no astrophysical nor cosmological problems associated with gravitino. 

The organization of the paper is as follows. In section 2, we briefly review the conformal 
gauge mediation scenario and discuss the spectrum of the SUSY standard model  (SSM)  sector in the case of 
strongly interacting conformal gauge mediation.
In section 3, we show explicit examples of the strongly interacting conformal gauge mediation. 
The last section is devoted to our conclusions with a further discussion. 

%%%%%%%%%%%%%%%%%%%%%%%%%%%
\section{Conformal Gauge-Mediation Scenario}
The conformal gauge-mediation scenario~\cite{Ibe:2007wp} is based on an extension of a dynamical SUSY breaking model, which is also a variant of the conformal SUSY breaking~\cite{Ibe:2008td}, 
where the SUSY breaking model is extended by introducing vector-like representations $(P,\bar P)$ 
as new flavors with the superpotential mass term
\begin{eqnarray}
W = \sum m P \bar P \ . \label{masst}
\end{eqnarray}
We choose the number of the new flavors so that the extended dynamical SUSY-breaking sector 
has a non-trivial infrared (IR)-fixed point in the massless limit of the new flavors ($m\to 0$).

The important assumption in the conformal SUSY breaking 
is that the extended SUSY-breaking model is in the vicinity of the IR-fixed point at the ultraviolet (UV)
 cut-off scale where we can neglect the mass of the new flavors.
Under this assumption, all the coupling constants in the SUSY-breaking sector immediately
converge to the values at the IR-fixed point once they evolve down to the IR from the UV cut-off scale.
Therefore, there remains no free parameter in the conformal SUSY-breaking sector 
at the IR scale.

At the far IR scale, the SUSY is broken dynamically after 
the conformal symmetry is broken at the decoupling scale (i.e. physical mass $ m_{\rm phys}$) of the new flavors,
\begin{eqnarray}
\label{eq:Pmass}
m_{\mathrm{phys}} = m \left(\frac{m}{M_{UV}}\right)^{\frac{\gamma_P}{1-\gamma_P}} \ ,
\end{eqnarray}
where $\gamma_{P}$ denotes the anomalous dimension of $P$ and $\bar P$ at the IR-fixed point, 
and $M_{\rm UV}$ is the scale of the UV cut-off \cite{Ibe:2007wp}.
Notice that since all the coupling constants of the SUSY-breaking sector are fixed on the IR-fixed point,
the relation between the mass term of the new flavors and the dynamical SUSY-breaking scale is uniquely determined, that is, the SUSY-breaking scale is related to the mass of the new flavors by
\begin{eqnarray}
\label{eq:decouple}
\Lambda_{\rm susy} \simeq c_{\rm susy}m_{\rm phys} \ ,
\end{eqnarray}
with a coefficient $c_{\rm susy}$.
Notice that the ratio $c_{\rm susy}$ is not a free parameter of the model but determined by the 
dynamics.
When the model is strongly interacting at the IR-fixed point above the mass scale $m_{\rm phys}$, the ratio $c_{\rm susy}$ is expected to be $O(1)$, since the gauge coupling constant of the SUSY-breaking sector blows up just below the scale of the decoupling of  the new flavors.

By embedding the SSM gauge group into the flavor symmetry of the new flavors,
they can serve as messenger particles. 
In this way, we can construct a model of conformal gauge mediation which possesses no tunable parameters except for the mass of the messengers, $m_{\rm phys}$.
Namely, in the conformal gauge mediation, all the soft masses are determined by the messenger mass $m_{\rm phys}$ as
\begin{eqnarray}
m_{\rm gaugino} &\simeq& \left( \frac{\alpha}{4\pi}\right)c_{\rm gaugino} m_{\rm phys}\ ,  \\
m_{\rm scalar} &\simeq& \left( \frac{\alpha}{4\pi}\right)c_{\rm scalar} m_{\rm phys}\ ,
\end{eqnarray}
with dimensionless coefficients $c_{\rm gaugino}\propto n_{\rm mess} c_{\rm susy}^{9}$ and 
$c_{\rm scalar}\propto n_{\rm mess}^{1/2} c_{\rm susy}^{3}$, where $n_{\rm mess}$ is the  number of the messengers.%
\footnote{
The mediation mechanism we are discussing here is similar to so-called
mediator model in Ref.\,\cite{Randall:1996zi} for $m_{\rm phys} \gg \L_{\rm SUSY}$.
As pointed out in Ref.\,\cite{ArkaniHamed:1998kj}, the $O(F)$ contribution to the gaugino mass 
is suppressed by higher-loops of the SSM gauge interactions compared 
to the $O(F)$ contribution to the scalar masses in the mediator model.
Here, $F$ denotes the $F$-term supersymmetry breaking in the SUSY breaking
sector.
On the other hand, the $O(F^{3})$ contributions starts at the one-loop 
diagram of the SSM (with higher loop diagrams of the SUSY breaking sector interactions).
Since we are interested in the model with $\sqrt{F}\sim m_{\rm phys}$,
the dominant contribution to the gaugino mass is not $O(F)$ but $O(F^{3})$.
}
We emphasize that the coefficients $c_{\rm gaugino}$ and $c_{\rm scalar}$ include no free parameters but have definite values depending on the 
model~\cite{Ibe:2007wp}.%
\footnote{
Here, we also assume that the $R$-symmetry is also broken spontaneously
at the scale of the order of $\Lambda_{\rm susy}$.
}

%%%%%%%%%%%%%%%
\subsection*{Mass estimation in strongly interacting models}
As we discussed in Ref.~\cite{Ibe:2007wp}, for $c_{\rm susy}\ll 1$, the gaugino mass 
is suppressed by about a factor of $n_{\rm mess}^{1/2}c_{\rm susy}^{6}$ than the sfermion masses.
Thus, if the model is weakly interacting at the IR-fixed point, the gaugino is much lighter than the sfermions.

When the model is strongly interacting at the IR-fixed point, however, the ratio $c_{\rm susy}$ can be $O(1)$.
In that case, we expect that the hierarchy between the gaugino and sfermion masses dissolves. The weak scale SUSY breaking without fine-tuning (i.e. without splitting SUSY spectrum) forces us to investigate the strongly interacting conformal gauge mediation.
In section~3, we will show explicit models of the conformal gauge mediation where the model is strongly interacting at the IR-fixed point.
Unfortunately, the precise prediction of soft masses is difficult in such cases since the 
messenger particles also take part in the strong interaction when they decouples.%
\footnote{Recently, generic properties of the gauge mediation associated with the strongly
interacting SUSY-breaking sector have been discussed in Ref.~\cite{Meade:2008wd}, although it is still difficult to obtain soft masses numerically.}
Here, instead, we estimate the gaugino and scalar masses as%
\footnote{For $c_{\rm susy}=O(1)$, the above approximation of the ratio, $m_{\rm gaugino}/m_{\rm scalar} \simeq n_{\rm mess}^{1/2} c_{\rm susy}^{6}$, breaks down.
}
\begin{eqnarray}
\label{eq:ssm}
m_{\rm gaugino} &\simeq&
 \frac{\alpha}{4\pi} n_{\rm mess}\Lambda_{\rm susy} \ ,\\ 
m_{\rm scalar}^{2} 
&\simeq &
\left(\frac{\alpha}{4\pi}\right)^{2} n_{\rm mess} \Lambda_{\rm susy}^{2} \ ,
\end{eqnarray}
 in the spirit of the naive dimensional analysis by assuming $c_{\rm susy}=O(1)$.%
 \footnote{
 The sign of the sfermion squared mass cannot be determined by perturbative analysis.
In this paper, we simply assume that the sfermions obtain positive mass squared.
 }

Notice that as the SUSY breaking scale is uniquely determined by the 
messenger mass (or equivalently by the SUSY breaking scale),
 the same holds for the gravitino mass,
\begin{eqnarray}
m_{3/2} = \frac{\Lambda_{\rm susy}}{\sqrt{3}M_{\rm PL}}  \ .
\end{eqnarray}
Here, $M_{\rm PL}\simeq 2.4\times 10^{18}$\,GeV denotes the reduced Planck scale.
Thus, in the conformal gauge mediation, there is a strict relation between the soft masses
in Eq.~(\ref{eq:ssm}) and the gravitino mass.
Interestingly, the relation predicts a very light gravitino in the case of the strongly interacting models, 
 That is, by requiring that the gaugino and the scalar masses are of the order of $1$\,TeV, 
we obtain
\begin{eqnarray}
\Lambda_{\rm susy} = O(10^{4-5})\,{\rm GeV} \ ,
\end{eqnarray}
which corresponds to the gravitino mass
\begin{eqnarray}
m_{3/2} \equiv \frac{\Lambda_{\rm susy}^{2}}{\sqrt{3} M_{\rm PL}} = O(0.01-1)\,{\rm eV} \ .
\end{eqnarray}
Therefore, we find that the conformal gauge mediation with no large hierarchy between
the gaugino and scalar masses predicts the gravitino mass $m_{3/2}\lsim O(1)$\,eV. 
Notice that such a small gravitino mass may be determined
 at the future collider experiments,
e.g. by measuring the branching ratio of the decay rate of the next to lightest  
superparticle~\cite{Hamaguchi:2007ge}.

From the cosmological point of view, the light gravitino of mass $m_{3/2} <O(10)$\,eV is very attractive 
since it shows no conflict with astrophysical and cosmological observations~\cite{Viel:2005qj}.
Moreover, as we will see in section~3, we can construct models with a stable SUSY breaking vacuum in our framework.
In such cases, the conformal gauge mediation model is quite successful in cosmology regardless of 
the detail of the thermal history of the universe.

So far, there have been some attempts to obtain models of gauge mediation 
with $m_{3/2}< O(10)$\,eV, where the SUSY breaking vacuum is stable
(see Refs.~\cite{Izawa:1997hu}-\cite{Izawa:2005yf} for example).
In those models, however, the motivation to choose the parameter to do so would still need to be explained. 
In the conformal gauge mediation, 
however,
 the prediction of the light gravitino is rather compulsory because there is no parameter to tune.

Before closing this section, we comment on another attractive feature of the conformal gauge mediation.
As briefly discussed in Ref.~\cite{Ibe:2007wp}, the messenger quarks are expected to be heavier than the messenger leptons by the QCD wave function renormalization effects to the messenger quarks.%
\footnote{In the usual gauge mediation models, this mechanism does not work, since the wave
function renormalization effects to the messenger masses are cancelled by the same 
effects to the coupling of the messengers to the SUSY-breaking field.}
Thus, the colored superparticles obtain relatively lighter masses 
compared with the usual gauge mediated SUSY breaking models, 
which makes superparticles more accessible 
at the Large Hadron Collider experiments than the usual gauge mediation models.

%%%%%%%%%%%%%%%%%%%%%%%%%%%%%
\section{Examples of Conformal Gauge Mediation}
In this section, we present two examples of the conformal gauge mediation 
where the ratio between the messenger scale and the SUSY breaking scale is expected 
to be $O(1)$, i.e. $c_{\rm susy}=O(1)$.
Although there are many choices for the dynamical SUSY-breaking which would be
extended to the conformal gauge mediation model,  we concentrate on the scenario in which the SUSY-breaking vacuum is stable with the consistent cosmology in mind.

The first example is a model based on the dynamical SUSY breaking of $SO(10)_h$
gauge theory with a spinor representation~\cite{Affleck:1984mf,Murayama:1995ng}.
According to a general procedure to realize conformal gauge mediation,
we add $N_f$ vector-like representation ${\bf 10}$ with a mass term in Eq.~(\ref{masst}).
For $7<N_{f}<21$, this model is known to have an non-trivial IR-fixed 
point~\cite{Pouliot:1996zh,Kawano:1996bd}.
As analyzed in Ref.~\cite{Ibe:2005qv,Kawano:2005nc}, 
the anomalous dimensions of the chiral superfields at the conformal fixed point can be computed by using the $a$-maximization technique~\cite{Intriligator:2003jj,Barnes:2004jj}:
\begin{eqnarray}
\gamma_{10} = \frac{-5 -24 N_f + N_f^{2} + \sqrt{ 2885 - N_{f}^{2}} }{-5 + N_f^{2}} \ .
\end{eqnarray}
For $N_{f} = 10$, we have $\gamma_{P}\simeq -0.97$.
Since the anomalous dimension of the messengers is close to the unitarity bound: $\gamma_P\simeq -1$,
the model is expected to be strongly interacting at the IR-fixed point, and hence, the ratio $c_{\rm susy}$
is expected to be $O(1)$.
By identifying subgroups of the flavor symmetry $SU(5)\subset SU(10)$ 
 with the gauge groups of the SSM, we obtain an example
of the strongly interacting conformal gauge mediation.%
\footnote{For the time being, we will neglect the effect of SSM gauge coupling to $\gamma_{P}$. At the very high energy scale, where the SSM gauge coupling constant could become large, this assumption might not be valid while we expect the deviation of the whole scenario from the picture presented here is small. We return to this point below.}

Another example is a model based on the dynamical SUSY breaking of 
$SU(5)_{h}$ with $\mathbf{10}+\bar{\mathbf{5}}$~\cite{Affleck:1983vc,Murayama:1995ng}. 
Again, we add $N_f$ vector-like quarks $\mathbf{5} + \bar{\mathbf{5}}$ (for $ 5<N_f <13$) to make 
the model have an non-trivial IR-fixed point~\cite{Ibe:2007wp} (see also \cite{Pouliot:1995me}).
We identify five out of $N_f$ flavors are messenger fields which are charged under the SSM gauge group.
The anomalous dimensions of the chiral superfields at the conformal fixed point can be computed by 
\begin{align}
\gamma_{\mathbf{5}} &= \gamma_{\bar{\mathbf{5}}} =
\frac{-85+8(-14+N_f)N_{f}+3\sqrt{5425-8N_{f}(1+N_f)}}{-25+8N_{f}(1+N_f)} \ .
\end{align}
For $N_f = 6$, we have $\gamma_P \simeq -0.82$, 
and hence, this model is also expected to have $c_{\rm susy}=O(1)$.
Thus, another example of the strongly interacting conformal gauge mediation model
is obtained  
by identifying the flavor symmetry $SU(5)\subset SU(6)$ with the gauge groups 
of the SSM.

\subsection*{Perturbative GUT?}
One unavoidable property of the strongly interacting conformal gauge mediation is the large beta function contribution to the SSM gauge coupling constant. This is due to the fact that the anomalous dimensions of the messengers will increase the number of messengers charged under the SSM gauge group.
The perturbativity of the standard-model gauge interactions demands that the number of the messengers $n_\mathrm{mess}$ should satisfy
\begin{eqnarray}
n_{\mathrm{mess}}  \lesssim \frac{150}{(1-\gamma_{P})\ln( {M_{\mathrm{GUT}}}/m_{\mathrm{phys}})}\ , \label{beta}
\end{eqnarray}
where we have included the higher loop effects of the SUSY-breaking sector through the anomalous
dimension $\gamma_{P}$ of $P$ and $\bar{P}$.
Here, we have used the NSVZ exact formula~\cite{Novikov:1983uc}--\cite{ArkaniHamed:1997mj} of the beta functions of the SSM gauge interactions. 
For $\gamma_{P}\simeq -1$ and $m_{\rm phys}=O(10^{5})$\,GeV
this condition is reduced to
\begin{eqnarray}
n_{\rm mess}  \lesssim 3 \ .
\end{eqnarray}
In the above two examples, the numbers of the messengers are $n_{\rm mess} = 10$ 
for the $SO(10)_h$ model and $n_{\rm mess} = 5$ for the $SU(5)_h$ model, respectively.%
\footnote{
In the model based on $SU(5)$, $N_{f}=5$ is enough to identify the flavor symmetry with
the SSM gauge group as discussed in Ref.~\cite{Izawa:2005yf}.
In this model, the SSM gauge couplings are expected not to blow up below the GUT scale,
since the SUSY breaking sector is asymptotically UV free,
although this model is not in the category of the conformal gauge mediation.
}  
Therefore, the standard model coupling constants blow up below the GUT scale  as long as the perturbative formula for the beta function \eqref{beta} is valid. 
However, this does not necessary mean that the theory is ill-defined above that scale: 
it is just a breakdown of the low-energy effective field theory description.
It rather suggests the presence of a dual description of the standard model at the high-energy scale,
where  the standard model itself can be realized as a weakly interacting dual gauge group
(we refer e.g. to \cite{Strassler} for an attempt).

Leaving the above interesting possibility aside, there are several possible ways to avoid the problem if we wish.
One way to recover the perturbative unification is to separate the messenger gauge group and 
the SSM subgroup of $SU(5)_{\rm GUT}$. 
For example, let us abandon identifying the subgroups of the flavor $SU(5)_{F}$ symmetry 
of the above $SU(5)_h$ 
SUSY breaking model with the SSM gauge group, and, instead, 
consider it as an independent gauge group.
Let us, then, assume that the flavor gauge symmetry $SU(5)_{F}$ 
and the SSM subgroups of $SU(5)_{\rm GUT}$ ($\supset SU(3)\times SU(2)\times U(1)$) 
break down to the diagonal subgroups, the (low-energy) SSM $SU(3)\times SU(2)\times U(1)$
at a scale $M_{5}$
by a VEV of a bi-fundamental field of the $SU(5)_{F}$ and the SSM gauge groups.
In this model, the messenger particles are charged under the low-energy SSM gauge groups
while they are neutral under the SSM gauge group above the scale $M_{5}$.
In this way, we can realize the above conformal gauge mediation model with $M_{UV}\lsim M_{5}$,
while the SSM gauge coupling constants do not receive large beta function contributions 
from the messengers above the scale $M_{5}$, which makes the perturbative GUT possible.%
\footnote{
The perturbative unification of the SSM gauge coupling is realized
when the following three conditions are satisfied.
1) $M_{5}$ is required to be close to $m_{\rm phys}$, 
so that the SSM gauge couplings do not receive large renormalization effects 
between $m_{\rm phys}$ and $M_{5}$.
2) The $SU(5)_{F}$ gauge theory is perturbative enough 
so that  the perturbativity condition of the SSM gauge couplings similar to Eq.\,(\ref{beta}) 
admits the newly introduced bi-fundamntal field.
3) The gauge coupling constant of $SU(5)_{F}$ is rather large at $M_{5}$, 
so that the gauge coupling constants of the SSM do not change so much at the threshold
scale $M_{5}$.
} 
We emphasize that although the breaking of $SU(5)_F\times SU(5)_{\rm GUT}$ to 
the diagonal subgroups introduces a new scale, the low energy physics is barely affected by the scale. 
Therefore, the philosophy of the conformal gauge mediation (i.e. unique low energy prediction) is still intact under this modification.
(See also the appendix A for the discussion of another possibility.)

%%%%%%%%%%%%%%%%%
\section{Conclusion and Discussion}
In this note, we have shown that the conformal gauge mediation admits the non-hierarchical 
SSM spectrum by considering a strongly interacting theory.
An interesting prediction of the strongly interacting conformal gauge mediation 
is the very light gravitino ($m_{3/2}<O(10)$\,eV), which is very attractive from 
a cosmological point of view.
As another attractive feature,  we can construct models with the stable
SUSY breaking vacuum.
In such models, there is no constraint on the thermal history of the universe
which is severely constrained if the vacuum is meta-stable.

Several comments are in order.
As we have discussed, the gravitino mass is predicted to be $O(1)$\,eV for strongly interacting conformal gauge mediation models.
In this case, the gravitino abundance cannot provide the mass density of the observed dark matter. 
Thus, there must be other candidates for the dark matter.
The most interesting candidate for the dark matter is the QCD 
axion~\cite{Weinberg:1977ma,Wilczek:1977pj}
which is involved in a solution to the strong CP-problem by the spontaneously breaking of the
anomalous Peccei-Quinn (PQ) symmetry~\cite{Peccei:1977hh} 
(with the breaking scale $f_{PQ}\simeq 10^{11}\,{\rm GeV}$~\cite{Turner:1985si}).
By assuming that the strong CP-problem is solved by the axion mechanism, we can picture
the SSM with $m_{3/2}<O(10)$\,eV, fully consistent with cosmology.

The introduction of the PQ-symmetry also provides us with an interesting perspective on the origin of the $\mu$-term.
With appropriate charge assignments for the PQ-breaking field (with a breaking scale $f_{PQ}$)
and the Higgs doublets under the PQ-symmetry, we can write down a higher dimensional term in the superpotential
\begin{eqnarray}
W = \frac{f_{\rm PQ}^{2}}{M_{\rm PL}} H_{u} H_{d} \ .
\end{eqnarray}
Thus, for $f_{PQ}\simeq 10^{11}$\,GeV, we obtain an appropriate size of for $\mu$-term, $\mu=O(1)\,$TeV, without causing another CP-problem.

We also comment on the dynamical tuning of the cosmological constant~\cite{Ibe:2008td}.
As discussed in Ref.~\cite{Ibe:2008td}, the dynamical tuning of the cosmological constant is 
realized in strongly interacting conformal SUSY breaking models 
for $\gamma_{P}\simeq-1$ and $M_{UV}\simeq M_{\rm PL}$,
by attributing the origin of the mass of the new flavors to the constant term in the 
superpotential.
The degree of the fine-tuning of the cosmological constant is greatly improved as a result 
of the dynamical tuning.
Since the conformal gauge mediation is based on the conformal SUSY breaking, 
it is an interesting question whether the strongly interacting conformal gauge mediation
can work with the dynamical tuning mechanism of the cosmological constant.

The immediate problem is that, as we have commented before, 
the conformal fixed point would be disturbed by the rather large SSM gauge coupling constants.
Thus, it is non-trivial whether the model admits $M_{UV} \simeq M_{\rm PL}$.
Having said that, the disturbance is expected to be significant only at very high energy scale 
(typically above the holomorphic Landau pole scale), 
and hence, there is a possibility that the conformal gauge mediation, as it stands, 
might work well with the dynamical tuning mechanism of the cosmological constant.

The model based on $SU(5)_h\times SU(5)_{F} \times SU(5)_{\rm GUT}$ gauge symmetry 
discussed at the end of section 3 may shed light on the other possibility.%
\footnote{See also the appendix A for another possibility. }
As we have discussed, the model admits the perturbative GUT unification of the SSM gauge couplings.
Thus, the effects of the SSM gauge coupling constants to the $SU(5)_h\times SU(5)_{F}$ sector
is not significant.
Now, let us go one step further and assume the gauge coupling constant of $SU(5)_{F}$
also has an IR-fixed point together with $SU(5)_h$.
In this case, we can extend the UV cut-off of the conformal phase from $M_{5}$ to $M_{\rm PL}$,%
\footnote{The anomalous dimensions of the bi-fundamental fields $\Phi$ (${\bf 5}_F\times \bar{\bf{5}}_{\mathrm{GUT}}$) and $\bar{\Phi}$ ($\bar{{\bf 5}}_F\times {\bf{5}}_{\mathrm{GUT}}$) that are charged under the standard model are $\gamma = -0.15$, so the perturbative GUT is achieved. We also note that the anomalous dimensions of the SUSY breaking sector are only slightly modified: the anomalous dimension of massive bi-fundamental fields $P$ (${\bf 5}_h\times\bar{\bf{5}}_{F}$) and $\bar{P}$ ($\bar{\bf 5}_h\times {\bf{5}}_{F}$) are given by $\gamma = -0.86$ for instance.}
which makes it possible to realize the dynamical tuning of the cosmological constant.

\section*{Acknowledgments}
The research of Y.~N. is supported in part by NSF grant PHY-0555662 and the UC Berkeley Center for Theoretical Physics.
The work of MI was supported by the U.S. Department of Energy under contract number 
DE-AC02-76SF00515.
This work was supported by World Premier International  Research Center Initiative (WPI Initiative), MEXT, Japan.

%%%%%%%%%%%%%%%%%
\appendix
\section{Cousin model of the conformal gauge mediation}
In this appendix, we consider a cousin model of the conformal gauge mediation 
where  the SSM spectrum has a strict relation with the gravitino mass in which
we again make use of the conformal SUSY breaking.

The model is based on the conformal SUSY breaking model of $SO(10)$ gauge group with a spinor representation. We introduce $N_f=10$ numbers of vector representation $P$ to make it conformal. In addition, for messengers, we add $SO(10)$ singlet superfield $X$ and $\bar{X}$ which are charged under $SU(5)_{\rm GUT}$ as $\bf 5$ and $\bar{\bf 5}$ respectively. The superpotential is given by
\begin{eqnarray}
W =  m PP + \frac{\lambda}{M_{\rm PL}} X\bar{X} PP \ .
\end{eqnarray}
We regard the mass term for $P$ as a small perturbation as before, but we assume that the quartic coupling $\lambda$ is in the vicinity the strongly interacting fixed point value (i.e. $\lambda_* \sim 1$).\footnote{This assumption is actually unnecessary for the phenomenological success of the model because, as we will see, the leading order spectrum does not depend on $\lambda$. We  here stick to the philosophy of the conformal gauge mediation, however.} 
Before turning on the mass deformation, the model is supposed to be in the conformal regime. The anomalous dimension of $P$ can be re-computed as $\gamma_{10} \simeq -0.97$ by using the $a$-maximization.

The conformal symmetry is broken by the mass term. 
As a consequence, the SUSY is dynamically broken at 
$m_{\rm phys}$ in Eq.~(\ref{eq:Pmass}) near the origin of singlet fields $X$, $\bar{X}$.
The effective dynamics of the messengers $X$ and $\bar{X}$ can be represented by the superpotential
\begin{eqnarray}
W_{\rm mess} = \frac{\lambda_*}{m_{\rm phys}} X\bar{X} P P \ , \label{mess}
\end{eqnarray}
due to the anomalous dimension of $X$, $\bar{X}$ and $P$.

We now set a dynamical assumption that the strong dynamics of the $SO(10)$ model would give VEV of $P$ as
\begin{eqnarray}
\label{eq:PPvac}
\langle PP \rangle \sim \Lambda_{\rm susy}^2 + \Lambda_{\rm susy}^3 \theta^2 \ , 
\end{eqnarray}
where $\Lambda_{\rm susy} \simeq m_{\rm phys}$. The messenger superpotential \eqref{mess} is, then,
\begin{eqnarray}
W_{\rm mess} = \lambda(\Lambda_{\rm susy} + \Lambda_{\rm susy}^2 \theta^2) X\bar{X} \ . 
\end{eqnarray}
At this stage, the effective dynamics of the model has been reduced to the conventional gauge mediation scenario. where we have $m_{\rm gaugino} \sim m_{\rm sfermion} \sim \alpha \Lambda_{\rm susy}/4\pi$ which are independent of the parameter $\lambda$.
Therefore,  the scale of the SSM spectrum is determined by 
only $\Lambda_{\rm susy}$ as in the conformal gauge mediation model.
Notice that, by the same argument we made in section 2, 
this model also predicts the light gravitino ($m_{3/2}<O(1)$\,eV).

An important feature of the cousin model is that 
the perturbativity of the SSM gauge couplings is intact up to the GUT scale.
Thus, we can easily justify the assumption that the UV cut-off scale of the conformal 
SUSY breaking sector to be the Planck scale, i.e. $M_{UV}\simeq M_{\rm PL}$.
Therefore, in this model, we can also realize the dynamical tuning of the cosmological 
constant~\cite{Ibe:2008td} by attributing the origin of the mass term of the new flavors 
in the conformal SUSY breaking sector.%
\footnote{This model has  a SUSY vacuum at $\vev{X}\sim \sqrt{w_{0}M_{\rm PL}}$,
and hence, the above vacuum in Eq.~(\ref{eq:PPvac}) is meta-stable.
Thus, a careful study is required to discuss the thermal history of the universe.}

\end{document}